\documentstyle[multicol,prl,aps,epsfig,eqsecnum]{revtex}
\begin{document}
\draft
\title{Measurement of the ($\gamma$,n) reaction rates of the
nuclides $^{190}$Pt, $^{192}$Pt, and $^{198}$Pt in the astrophysical
$\gamma$ process}
\author{
 K.~Vogt,$^1$ P.~Mohr,$^1$ M.~Babilon,$^1$ J.~Enders,$^1$ 
  T.~Hartmann,$^1$ C.~Hutter,$^1$ T.~Rauscher,$^{2,3}$ S.~Volz,$^1$ 
  A.~Zilges$^1$
}
\address{$^1$
  Institut f\"ur Kernphysik, Technische Universit\"at Darmstadt,
  Schlossgartenstrasse 9, D-64289 Darmstadt, Germany
}
\address{$^2$
  Institut f\"ur Physik, Universit\"at Basel, 
  Klingelbergstrasse 82, CH-4056 Basel, Switzerland
}
\address{$^3$
  Department of Astronomy and Astrophysics, UCSC,
  Santa Cruz, CA 95064, USA
}
\date{\today}
\maketitle
\begin{abstract}
The nucleosynthesis of heavy neutron-deficient nuclei in a stellar
photon bath at the temperatures relevant for the astrophysical
$\gamma$ process was investigated. In order to derive ($\gamma$,n)
cross sections and reaction rates, the stellar photon bath was
simulated by the superposition of several bremsstrahlung spectra with
different endpoint energies. As a first test for this method, the
($\gamma$,n) reaction rates of the platinum isotopes  $^{190}$Pt,
$^{192}$Pt, and $^{198}$Pt were derived.
The results are compared to other experimental data and 
theoretical calculations.
\end{abstract}



\begin{multicols}{2}
\narrowtext

\section{Introduction}
\label{sec:intro}

The bulk of the nuclei heavier than iron have been synthesized by
neutron capture in the astrophysical $r$ and $s$ processes. Those
neutron-capture processes cannot account for the synthesis of some of
the heavy ({\it A} $\ge$ 100) neutron-deficient nuclei. These nuclei
are shielded from the chain of $\beta^-$ decays by other stable
isobars. The production mechanism for these so-called p-nuclei is
photodisintegration in the astrophysical $\gamma$ process by
successive ($\gamma$,n), ($\gamma$,p), and ($\gamma$,$\alpha$)
reactions. The natural abundances for the p-nuclei are very low in the
order of 0.01 \% to 1 \%. A complete list of the p-nuclei can be found
in Table 1 of Ref.~\cite{Lam92}. The starting point for the
photon-induced reactions are heavy seed nuclei which have been
synthesized in the $r$ and $s$ processes.

In order to reproduce the abundances of p-nuclei
measured in the solar system, temperatures in the $\gamma$ process
must be in the range of $T_9 = 2-3 $
($T_9$ is the temperature in 10$^9$ K),
densities about $\rho \approx 10^6 $ g/cm$^3$, and time scales
$\tau$ of the order of seconds.  A possible astrophysical
site which fulfills these requirements
could be the oxygen- and neon-rich layers of type II
supernovae. However, there has been no definite conclusion to this
question yet. Details about the $\gamma$ process and its astrophysical
scenarios can be found
in the reviews by Lambert~\cite{Lam92}, Arnould and
Takahashi~\cite{Arn99}, Langanke~\cite{Lan99}, Wallerstein {\it et
al.}~\cite{Wal97} and in
Refs.~\cite{Ito61,Woo78,Ray90,Ray95,Pra90,How91}.

For the calculation of the p-nuclei abundances resulting from the
$\gamma$ process, large reaction networks containing all relevant
nuclei and reaction rates are needed \cite{Ray90}. Until now, there
have been almost no experimental data available for these reaction
rates in the relevant energy region.  Although in the last decades a
large number of ($\gamma$,n) cross sections have been measured around
the giant dipole resonance (GDR), the energies of astrophysical
interest are much lower, and practically no data exist for the
p-nuclei because of their low abundance.  All reaction rates have been
derived theoretically, using statistical model calculations.  Reliable
experimental data would be a great improvement to reduce the nuclear
physics uncertainties of astrophysical model calculations, especially
because of the typical uncertainties of such statistical model
calculations which are at least of the order of a factor of 2.

As a starting point for the measurement of the needed reaction rates,
we have measured the ($\gamma$,n) reaction rates of the platinum
isotopes $^{190}$Pt(natural abundance: 0.014\%), $^{192}$Pt(0.782\%), and
$^{198}$Pt(7.163\%). 
The natural abundances have been taken from~\cite{Ros98}.\\
The ($\gamma$,n) reaction rate for a nucleus in a thermal
photon bath is given by
\begin{equation}
\lambda(T) =
  \int_0^\infty 
  c \,\, n_\gamma(E,T) \,\, \sigma_{(\gamma,{\rm{n}})}(E) \,\, dE
\label{eq:gamow}
\end{equation}
with the speed of light $c$ and the cross section of the ($\gamma$,n)
reaction $\sigma_{(\gamma,n)}$.  The number of photons $n_\gamma (E,T)$ 
at energy $E$ per unit of volume and per energy interval is given by the 
well-known Planck distribution
\begin{equation}
n_\gamma(E,T) = 
  \left( \frac{1}{\pi} \right)^2 \,
  \left( \frac{1}{\hbar c} \right)^3 \,
  \frac{E^2}{\exp{(E/kT)} - 1}
\label{eq:planck}
\end{equation}
For the measurement of the ($\gamma$,n) reaction rates we used the
method of photoactivation. We irradiated platinum samples with 
bremsstrahlung. 
Then we measured the number of decays of generated unstable platinum 
nuclei $^{189,191,197}$Pt.

The biggest difficulty in the determination of the ($\gamma$,n)
reaction rates is the reproduction of the thermal photon bath. In our
experiment, the platinum samples were irradiated with bremsstrahlung.
We are able to generate a quasi-thermal photon spectrum in
the relevant energy region by the superposition of several
bremsstrahlung spectra with different endpoint energies.
This idea has been presented in our previous paper~\cite{plb00}.

\section{Experimental setup}

The irradiation of the platinum samples was performed at the real
photon facility of the superconducting Darmstadt linear electron
accelerator S--DALINAC~\cite{Ric96,Mohr99,Zil00}. This setup is mainly
used for Nuclear Resonance Fluorescence experiments which can be
performed up to endpoint energies of 10 MeV without disturbing
neutron-induced background~\cite{Mohr99}.

As targets we used metallic platinum disks of natural isotopic
composition with a dia\-meter of 2{\it r} = 20 mm and a thickness of
{\it d} = 0.125 mm with masses of around 800 mg.  The platinum disks
were sandwiched between two thin boron layers with masses of about 650
mg each.  In order to normalize the photon flux we measured spectra of
resonantly scattered photons from nuclear levels of $^{11}$B during
the activation.  For this measurement, we used two high purity
germanium (HPGe) detectors with 100 \% efficiency (relative to a 3''
$\times$ 3'' NaI detector) which were mounted at $90^\circ$
and $127^\circ$ relative to the beam axis of the incoming photons.
Further information on the ($\gamma$,$\gamma'$) experiments can be
found in \cite{Hart00}.

\begin{figure}
\epsfig{figure=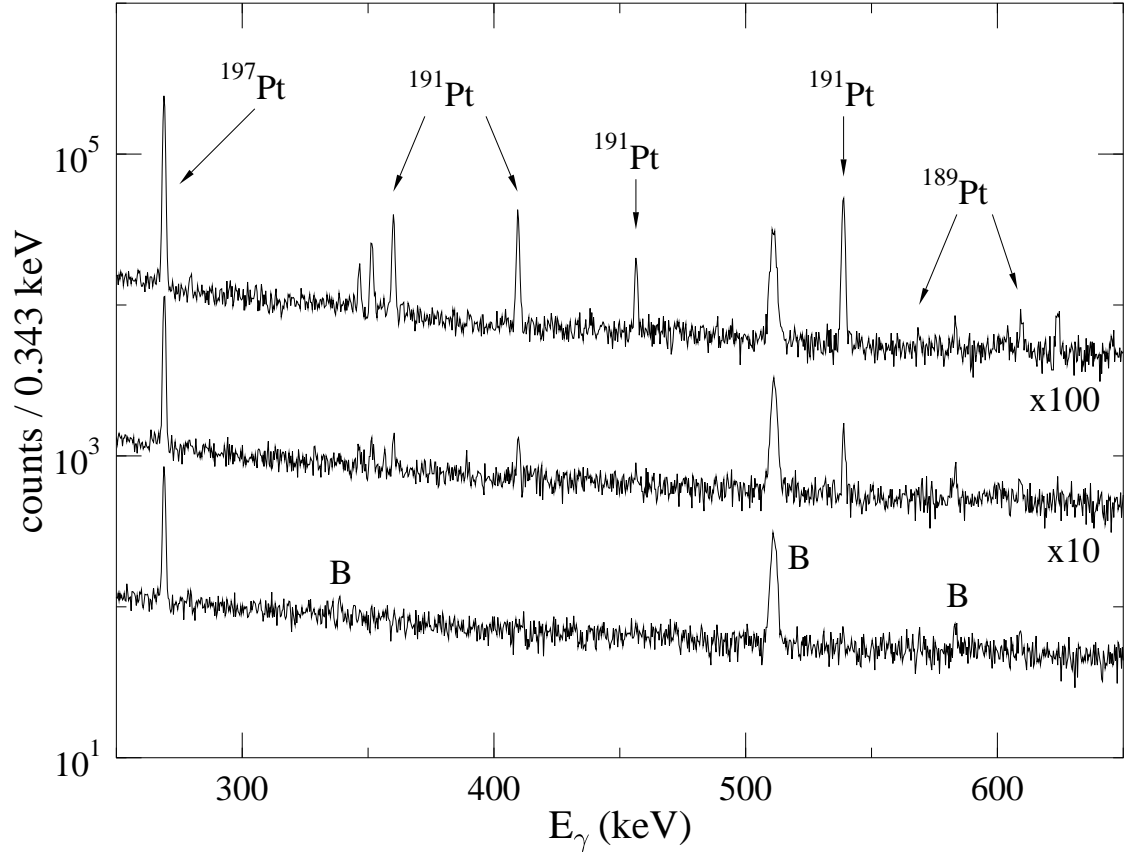,bbllx=16,bblly=97,bburx=556,bbury=509,
width=15.0cm}
\caption{\label{fig:spec_act} 
  Photon spectra of the activated platinum disks at the endpoint
  energies of $E_0 = 9900\,(\times 100),\,9450\,(\times 10),$ and $9000$\,keV, from top to bottom.
  Shown is the energy region between 250 and 650 keV.
  For a full spectrum with all relevant lines, see\protect\cite{plb00}.
  The main peaks from the decay of
  $^{189}$Pt,
  $^{191}$Pt, and
  $^{197}$Pt
  are indicated by arrows. Additional peaks from the backgound are
  labelled with B. The decay lines of $^{189}$Pt from the
  $^{190}$Pt($\gamma$,n)$^{189}$Pt reaction are close to the
  sensitivity limit of this experiment because of the low 0.014\%
  natural abundance of $^{190}$Pt. The lines  of $^{191}$Pt from the
  $^{192}$Pt($\gamma$,n)$^{191}$Pt reaction can hardly be seen 
  in the lowest spectrum because the endpoint energy of $E_0 = 9000 $\,keV
  is close to the neutron separation 
  energy of 8676 keV (see Table~\protect\ref{tab:platin}).
}
\end{figure}

 The platinum samples were irradiated for about 24 h, then their activity
 was measured for another 24 h.  For the measurement of the activity
 the samples were mounted directly in front of a third HPGe detector
 with 30 \% relative efficiency. Some typical activation spectra are 
 shown in Fig.~\ref{fig:spec_act}. Altogether, six measurements with
 endpoint energies of the bremsstrahlung spectra between $7650-9900$
 keV in steps of 450 keV were performed.

\section{Analysis of the experimental data}
\label{sec:analysis}

The aim of this experiment was the determination of the ($\gamma$,n)
reaction rates of several platinum isotopes. These reaction rates are
given by Eq.~(\ref{eq:gamow}).

The result of the analysis of the platinum activation spectra is an
integral over the ($\gamma$,n) cross section:
\begin{equation}
I_\sigma =
  \int_0^\infty 
   N_\gamma^{\rm{brems}}(E_{\rm{0}},E) \,\, \sigma_{(\gamma,{\rm{n}})}(E) \,\, dE
\label{eq:integral}
\end{equation}
where $N_\gamma^{\rm{brems}}(E_{\rm{0}},E)$ is the total number of
bremsstrahlung photons at energy {\it E} per area and per energy
interval during the irradiation.  $E_{\rm{0}}$ is the endpoint energy
of the respective bremsstrahlung spectrum. In order to determine the
($\gamma$,n) reaction rates from these results we used two different
methods (see Sect.~\ref{sec:convent}, \ref{sec:superposition}).

\subsection{The Gamow-like window for ($\gamma$,n) reactions}
\label{subsec:gamow}
\begin{figure}
\epsfig{figure=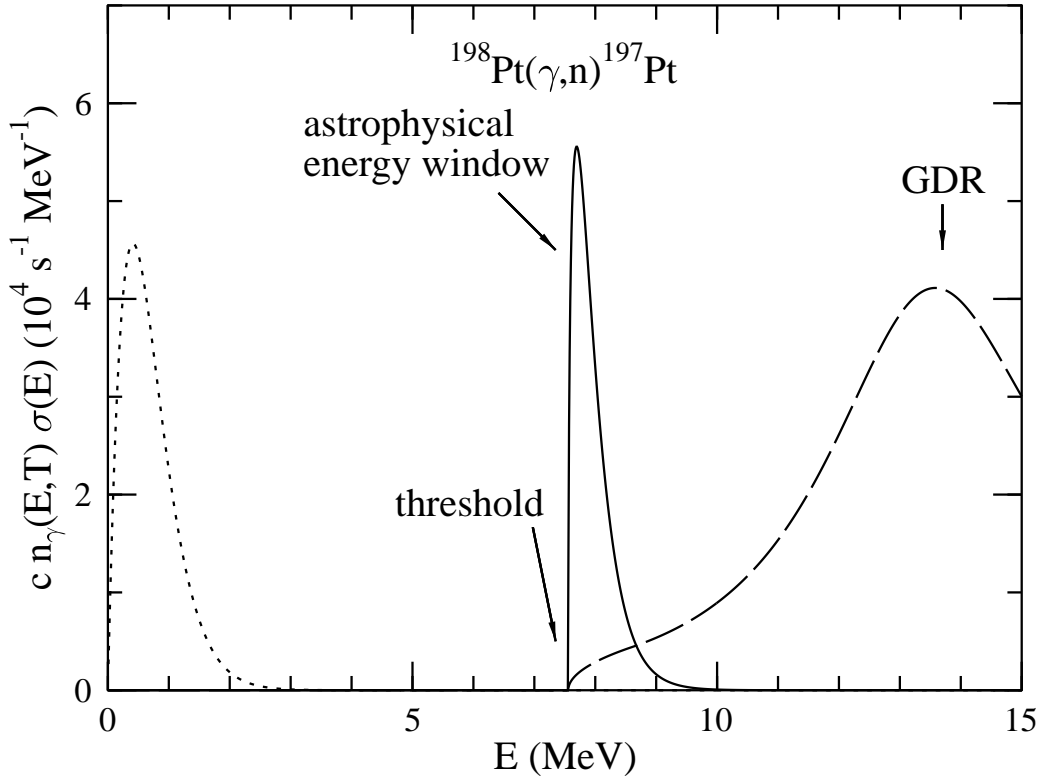,bbllx=100,bblly=90,bburx=380,bbury=290,
width=15.0cm}
\caption{\label{fig:gamow} Relevant energy window for ($\gamma$,n)
reactions in a thermal photon bath with the temperature $T_9 = 3.0$.
The integrand of Eq.~(\protect\ref{eq:gamow}) is given by the thermal
Planck distribution $n_\gamma(E,T)$ (dotted line) times the
($\gamma$,n) cross section $\sigma(E)$ (dashed line).
Note that the maximum of the integrand is located about $kt/2$ 
above the threshold energy of $E_{\rm{thr}} =
7557$\,keV, which was taken for the 
$^{198}$Pt($\gamma$,n)$^{197}$Pt reaction from \protect\cite{Audi95}. 
The GDR parameters were taken
from experimental data by Goryachev {\it et al.}\protect\cite{Gor78}, and
the threshold behavior $\sigma \sim \sqrt{E-E_{\rm{thr}}}$ was matched
to the Lorentzian shaped cross section of the GDR 1\,MeV above the
threshold. Note that the assumption of a typical threshold behavior is
not necessary for the determination of the quasi-thermal decay rate
$\lambda_{\rm{qt}}$ (see text).  }
\end{figure}

The integrand of Eq.~(\ref{eq:gamow}) is given by the product of the
photon flux $c\,n_\gamma$, which decreases exponentially with
increasing energy, and the ($\gamma$,n) cross section
$\sigma_{(\gamma,n)}$, which increases with {\it E} to the 
Giant Dipole Resonance (GDR).
Additionally, the threshold behavior has to be parametrized because
the Lorentzian parametrization of the GDR is not valid close to the
($\gamma$,n) threshold (see Sect.~\ref{sec:convent}).
The Planck distribution, a typical ($\gamma$,n) cross section, and the
resulting integrand of Eq.~(\ref{eq:gamow}) are shown in 
Fig.~\ref{fig:gamow} for $^{198}$Pt at $T_9 = 3.0$.
The maximum of the integrand is located at about {\it kT/2} above the
($\gamma$,n) threshold.
The behavior of the integrand is similar to the
well-known Gamow window (see e.g.\ Ref.~\cite{rolfs}) for 
charged particle reactions at
thermonuclear energies. The properties of the window for ($\gamma$,n)
reactions have been discussed in detail in~\cite{nuinco}.

\subsection{The conventional analysis}
\label{sec:convent}
One can assume the shape of the ($\gamma$,n) cross section to show a
typical threshold behavior:
\begin{equation} 
\sigma(E) = \sigma_0 \cdot \sqrt{(E-E_{\rm{thr}})/E_{\rm{thr}}} \quad \quad.
\label{eq:sqrt}
\end{equation}
This equation holds only in the vicinity of the reaction threshold.
Nevertheless, it should be sufficiently accurate since only a small
energy region above the threshold energy is relevant for the analysis.
By combining Eqs.~(\ref{eq:integral}) and (\ref{eq:sqrt}) it is
possible to derive the parameter $\sigma_0$ from our experimental
data. From the  parameter $\sigma_0$ we calculated the ($\gamma$,n)
reaction rate using Eq.~(\ref{eq:gamow}).
The serious drawback of this method 
is that it is not possible to estimate 
how much the real shape of $\sigma$(E) deviates from the approximation
 by Eq.~(\ref{eq:sqrt}).

\subsection{The superposition of bremsstrahlung spectra}
\label{sec:superposition}

\begin{figure}
\epsfig{figure=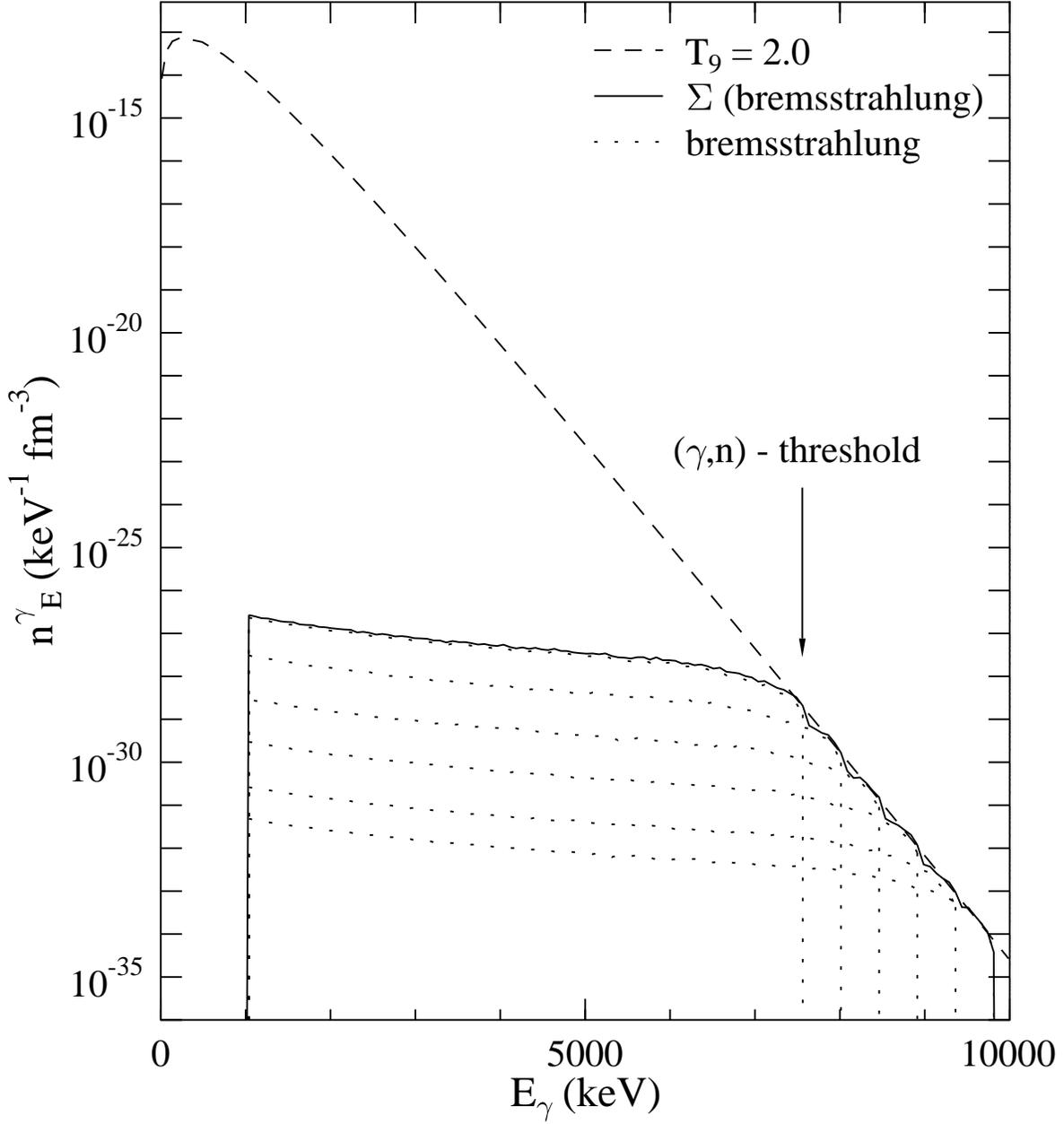,bbllx=35,bblly=86,bburx=542,bbury=657,
width=15.0cm}
\caption{\label{fig:super} The superposition of several
brems\-strahlung spectra (full line) with different endpoint energies
$E_0$ is compared to the thermal Planck spectrum $n_\gamma(E,T)$
(dashed line) at the temperature of $T_9 = 2.0$.  Good agreement is
found from 7.5 to 10\,MeV with the superposition of only six endpoint
energies. The six contributing brems\-strahlung spectra
$N^{\rm{brems}}_\gamma (E_{0,{\rm{i}}},E)$ are shown as dotted
lines. The arrow indicates the lowest ($\gamma$,n) threshold in our
experiment from the $^{198}$Pt($\gamma$,n)$^{197}$Pt reaction
($E_{\rm{thr}} = 7557$ keV). A figure which shows the superposition
for a temperature of $T_9 = 2.5$ can be found in\protect\cite{plb00}. }
\end{figure}

In order to derive the ($\gamma$,n) reaction rates directly from our
experimental data, that is to say without any assumptions on 
the shape of
the $\sigma(E)$ curve, we approximated the thermal Planck spectrum
$n_\gamma(E,T)$ in Eq.~(\ref{eq:planck}) by the superposition of
several bremstrahlung spectra with different endpoint energies:
\begin{equation}
  c \, n_\gamma(E,T) \approx \sum_i a_i(T) \, \cdot \,
  N_\gamma^{\rm{brems}}(E_{\rm{0,i}},E)
\label{eq:approx}
\end{equation}
where the $a_i(T)$ are strength coefficients which have to be adjusted
for each temperature {\it T}. With these strength coefficients, the
($\gamma$,n) reaction rates can be obtained by combining
Eq.~(\ref{eq:approx}) with Eq.~(\ref{eq:gamow}):
\begin{equation}
\lambda(T) = \sum^N_{i=1} a_i(T) \, \cdot \, \int
N_\gamma^{\rm{brems}}(E_{\rm{0,i}},E) \, \, \sigma (E) dE
\end{equation}
An example for this superposition is shown in Fig.~\ref{fig:super}.
In the energy range from 7.5 to 10 MeV the agreement between the
thermal Planck spectrum at $T_9 = 2.0$ and the weighted sum of the 
different bremsstrahlung spectra is reasonably good.  
Typical deviations are of the order of 10\%. Obviously, 
the huge deviations at energies below the ($\gamma$,n) 
reaction threshold are not relevant for our analysis.

\subsection{Analysis of the platinum activation spectra}
\label{subsec:ptact}
Fig.~\ref{fig:spec_act} shows three partial
platinum activation spectra.
Lines from the decay of the platinum isotopes $^{189}$Pt, $^{191}$Pt,
and $^{197}$Pt can clearly be identified in the upper spectrum
which shows the decay of a platinum sample which has been irradiated 
with a bremsstrahlung spectrum with an endpoint energy of 9900 keV. 
In the two lower spectra, which correspond to 
endpoint energies of 9450 keV and 9000 keV, respectively, the lines from 
the decay of $^{189}$Pt and $^{191}$Pt 
vanish because the endpoint energies of the bremsstrahlung
get close to the neutron separation 
energies of these isotopes (see Table~\ref{tab:platin}).
In the complete activation spectrum (see~\cite{plb00}) several additional lines, 
including two lines from the decay of $^{195m}$Pt can be identified; 
this isomer is mainly
populated by the ($\gamma$,$\gamma'$) reaction.  
The spectra were analyzed using the computer code tv~\cite{TV}.\\
For the derivation of
the ($\gamma$,n) reaction rates of the platinum isotopes $^{190}$Pt,
$^{192}$Pt and $^{198}$Pt, the  integral~(\ref{eq:integral}) must
be calculated from the number of counts {\it A} in the decay lines of the
platinum isotopes $^{189}$Pt, $^{191}$Pt and $^{197}$Pt, respectively.
The dependence of these quantities is given by
\begin{equation}
\label{equ:bla}
A = \epsilon \cdot I_\gamma \cdot \frac{T_{\rm{life}}}{T_{\rm{real}}}
\cdot \frac{N_{\rm{decay}}}{N_{\rm{total}}} \cdot N_{\rm{Pt}} \cdot
\int \sigma(E) \, N_\gamma^{\rm{brems}} \, dE
\end{equation}
where $\epsilon$ is the absolute detector efficiency, $I_\gamma$ the
absolute intensity of the platinum decay lines per decay, 
$T_{\rm{life}}$/$T_{\rm{real}}$ the ratio from
lifetime to realtime, and $N_{\rm{Pt}}$ the total number of platinum
nuclei of the respective isotope in the target. 
$N_{\rm{decay}}$/$N_{\rm{total}}$  is the ratio between the number
of nuclei which decay during the measurement and the total number
of produced unstable nuclei.
This ratio is given by:
\begin{equation}
 N_{\rm{decay}} / N_{\rm{total}} = e^{-\lambda
 T_{\rm{loss}}} \cdot \frac{(1 -  e^{-\lambda
 T_{\rm{irr}}})}{\lambda \, T_{\rm{irr}}}
 \cdot (1 - e^{-\lambda T_{\rm{measure}}})
\end{equation}
where $T_{\rm{measure}}$ is the duration of the measurement,
$T_{\rm{irr}}$ the duration of the irradiation, and $T_{\rm{loss}}$ the
time between the end of the irradiation and the beginning of the
measurement. 
These equations only hold for a constant production rate during the
irradiation, which is approximately given for our experiment since the
electron beam current was approximately constant.

The energies and relative intensities of the examined lines of the
platinum isotopes are shown in Table~\ref{tab:platin}. Note that the
relatively big uncertainties in the absolute intensities are
responsible for the major part of the errors in our results as seen in
Table~\ref{tab:summ}.

For the calculation of the factor $N_{\rm{decay}} / N_{\rm{total}}$,
precise knowledge of the half-lives of the produced unstable
platinum isotopes is necessary. Because of that, those half-lives
have been determined in an additional measurement~\cite{mohr2000}.

\subsection{Determination of the photon flux}
\label{subsec:flux}
For the approximation of the thermal Planck spectrum the precise shape
of the different bremsstrahlung spectra especially in the high energy
region had to be determined.  Therefore Monte Carlo simulations using
the computer code GEANT~\cite{GEANT} were performed. To check the
uncertainties of the GEANT calculations ($\gamma$,$\gamma'$)
measurements of well known $^{11}$B lines have been performed during
the irradiation of the platinum samples in order to normalize the
GEANT generated photon spectra.  At the energies of these $^{11}$B
lines, the absolute photon flux could be calculated.  For these
calculations, the efficiencies of the two detectors used for the 
measurement during the
irradiation of the platinum samples had to be calibrated up to the
energy of 10 MeV (see Sect.~\ref{subsec:eff}).

However, the shape of the bremsstrahlung spectra 
derived from the results of the
($\gamma$,$\gamma'$) measurements
and the spectra from the GEANT calculations show deviations near
the  endpoint energy. 
Unfortunately, the shape of the bremsstrahlung spectra could not be
determined by the $^{11}$B lines alone, because there are only two lines
(see Tab.~\ref{tab:bordaten}) in the high energy region.

Therefore, we tried to reproduce the measured $^{11}$B lines by using
theoretical formulas. The energy distribution of bremsstrahlung is
approximately given by the Schiff formula~\cite{Sch51}.  However, this
formula is only valid for bremsstrahlung from
thin radiator targets, which means that each
electron participates only in one scattering process and therefore
emits only one photon.  In our experiment, the electron beam was
completely stopped within the radiator target.  
Thus, neither the Schiff formula
nor newer approximation formulas given by Seltzer and
Berger~\cite{sel85} reproduce the shape of our bremsstrahlung spectra
with sufficient accuracy.  Especially in the high energy region, the
deviations are still considerable.

\begin{figure}
\epsfig{figure=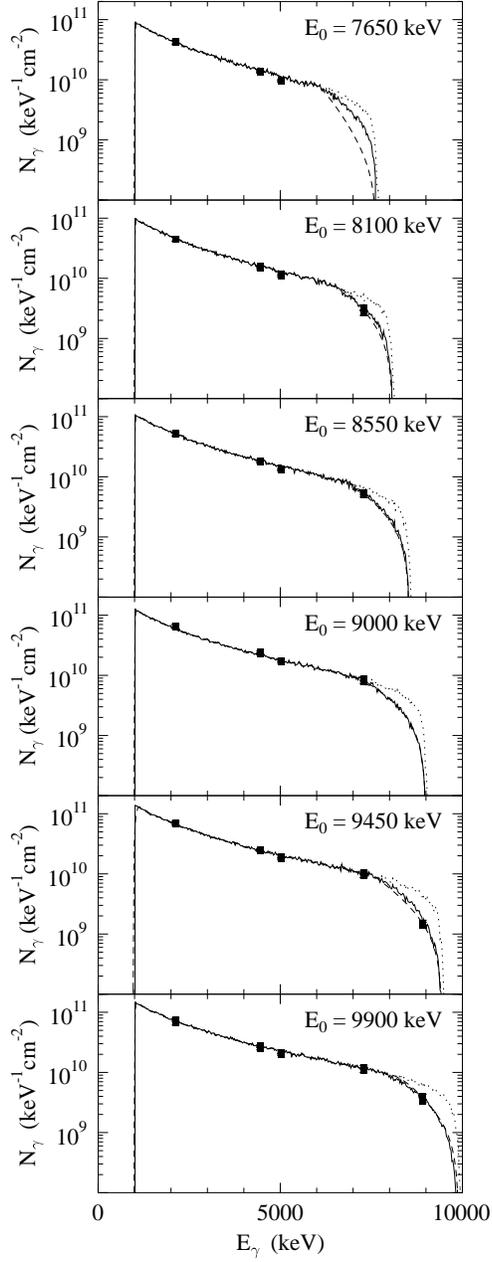,bbllx=0,bblly=4,bburx=318,bbury=840,
width=6.5cm}
\caption{\label{fig:borspektren} Photon flux spectra calculated using
the computer code GEANT (dotted line) at different endpoint energies
$E_0$.  In the high energy region, the GEANT spectra have been
adjusted by the correction factor $F$ in Eq.~(\protect\ref{eq:faktor})
(full line). The dashed lines
show the results of the cubic spline interpolation. The squares
represent the data points from the $^{11}$B($\gamma$,$\gamma'$)
measurement.  }
\end{figure}

Therefore, the shape of the bremsstrahlung spectra was interpolated in
the high energy region between 80 \% and 100 \% of the endpoint energy
using cubic
splines with the $^{11}$B lines as points of support.  Unfortunately,
in the spectrum with the lowest endpoint energy of 7650 keV, no
$^{11}$B line can be seen in the relevant energy region. Therefore, 
no points of support are 
available for the spline interpolation of this spectrum.  So we adjusted
the GEANT calculated spectra in the high energy region by multiplying
an energy-dependent correction factor {\it F}:
\begin{equation}
\label{eq:faktor}
F = \left\{ \begin{array}{c@{\quad;\quad}l}
1 - 5 \cdot 10^{-4} \cdot \left(\frac{E - 0.8 E_{\rm{0}}}{\rm{keV}}
\right)^{3/4} & E > 0.8\,E_0\\
1 &  E \le 0.8\,E_0 
\end{array} \right.
\end{equation}
where $E_0$ is the endpoint energy of the respective bremsstrahlung
spectra.  The shapes of the bremsstrahlung spectra resulting from both
methods are shown in Fig.~\ref{fig:borspektren} for all measured
endpoint energies. The results of the spline interpolation are
represented by the dashed lines, the results of the GEANT simulations
multiplied by the correction factor $F$ are represented by the straight
lines.  The good overall agreement between these curves and the data
points of the $^{11}$B measurements is quite satisfactory.
Since the spline interpolation is not reliable for the spectrum with 
7650 keV endpoint energy, we used the bremsstrahlung spectra 
adjusted by the
correction factor $F$ for the calculations of the reaction rates.
The correction factor $F$ leads to a significant reduction of the 
photon flux close to the endpoint energy $E_0$. 
A precise knowledge of this energy region is essential for 
the superposition of the quasi-thermal spectrum. 
If one neglects this correction in the bremsstrahlung spectra,
the reaction rates are underestimated by up to a factor of 2.

\subsection{Determination of detector efficiencies}
\label{subsec:eff}

\subsubsection{Detectors for photon flux normalization}
\label{subsubsec:detflux}
The calibration measurement was performed in two steps.
First, a measurement of the absolute efficiency
using $^{60}$Co, $^{137}$Cs, and $^{22}$Na calibration sources was
performed. This was followed by a second measurement using a $^{56}$Co source.
Since the intensity of this source was not calibrated, only a
relative efficiency could be determined, but these data points were
adjusted to fit those of the absolute efficiency measurement. 
Since there were
no decay lines of these sources at energies above 3548.3 keV, an
additional GEANT calculation was performed to determine the
efficiency in the high energy region. 

\begin{figure}
\epsfig{figure=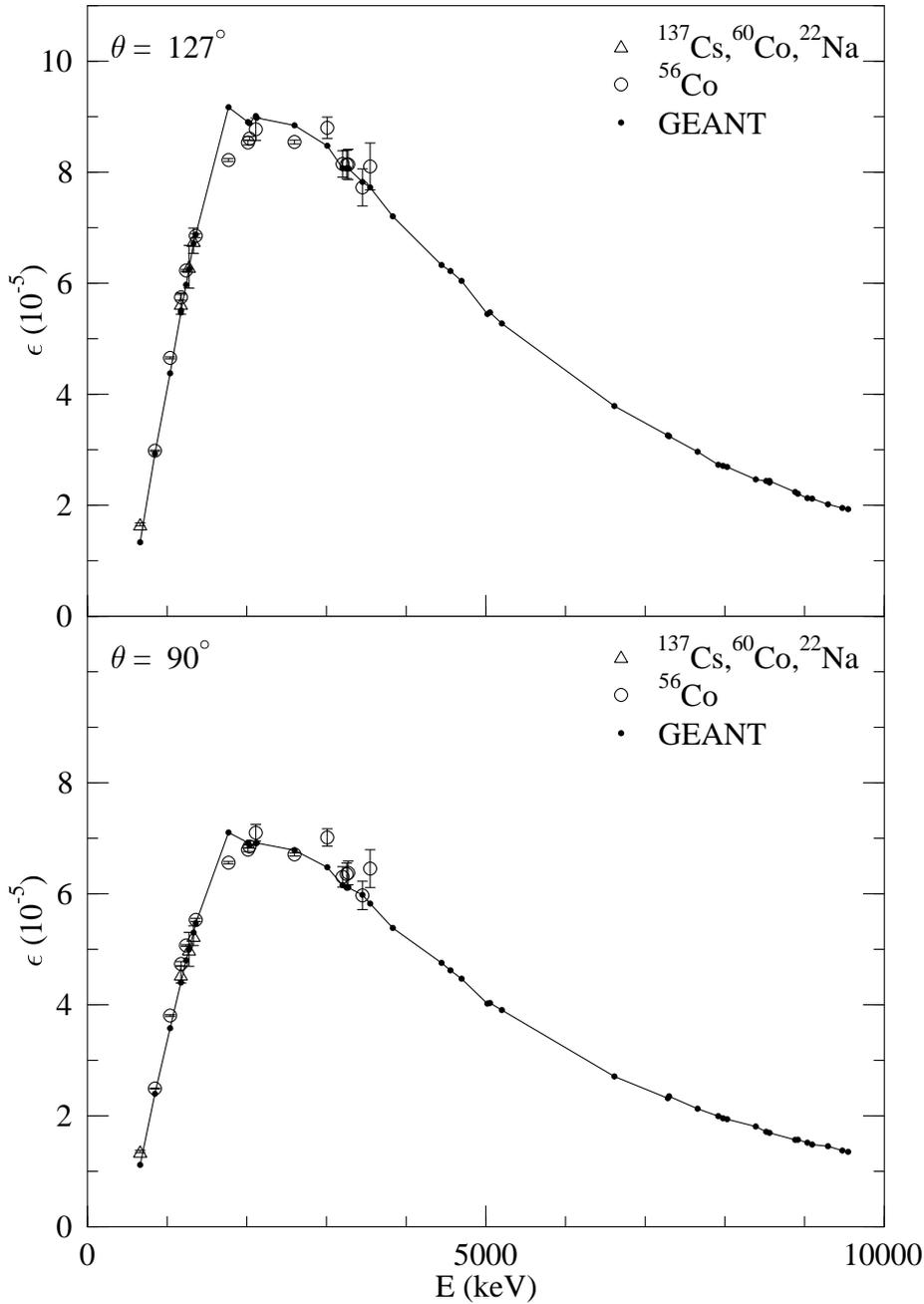,bbllx=20,bblly=40,bburx=500,bbury=750,
width=12.0cm}
\caption{
  \label{fig:abseff} 
Absolute efficiencies of the two detectors used for the
$^{11}$B($\gamma$,$\gamma'$) measurements.
The detectors were placed at $\theta = 90^\circ$ and $127^\circ$
relative to the incoming photon beam at distances of about 25 cm.
The efficiencies have been determined using several calibration
sources and are compared to a GEANT simulation.  
The data points from the GEANT simulation  are connected 
by a line to guide the eye. The decrease at low
energies comes from lead and copper filters in front of the detectors
to reduce the background at low energies.
}
\end{figure}

As can be seen in Fig.~\ref{fig:abseff}, the results
of the measurement and the calculation correspond well with each other,
and it has been shown that the detection efficiencies of large volume
HPGe detectors can be calculated with good accuracy \cite{Koe99}.
The decrease in the efficiency at low energies results from lead and
copper filters which were mounted in front of the detectors in order to
reduce low energy background.

\subsubsection{Detector for activation measurements}
\label{subsubsec:detact}
\begin{figure}
\epsfig{figure=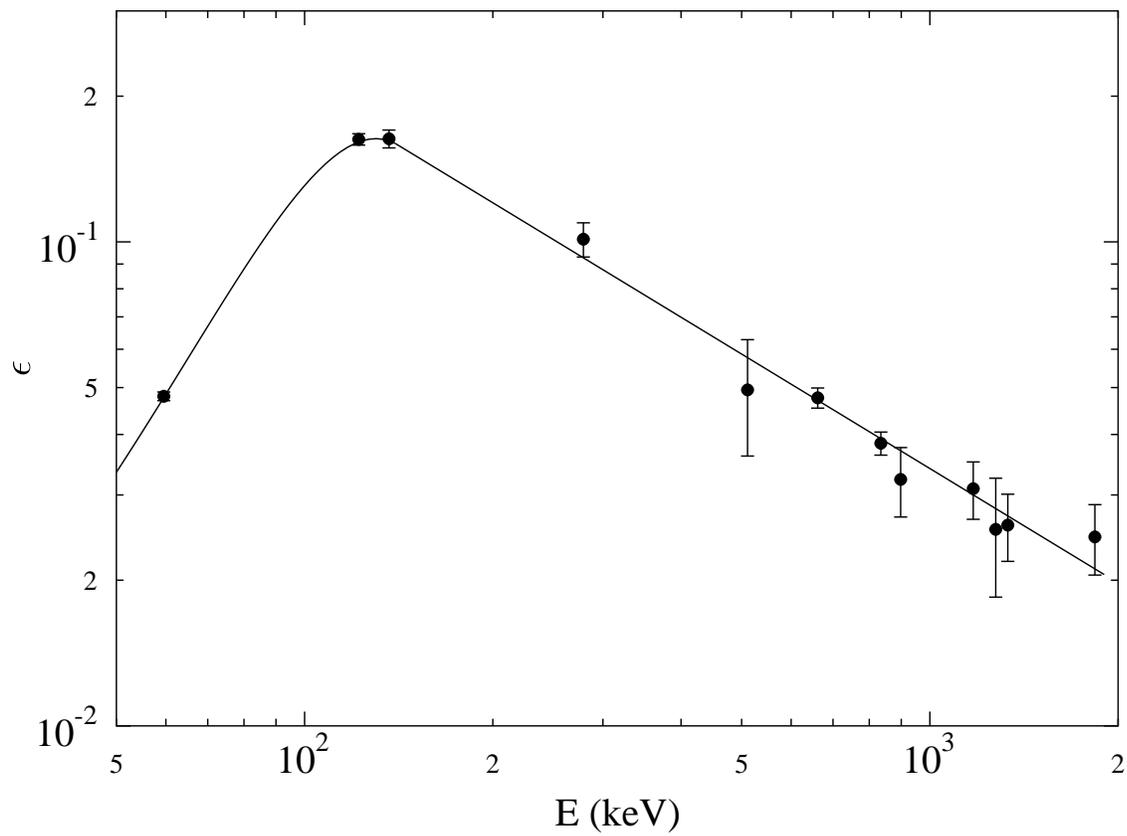,bbllx=53,bblly=123,bburx=500,bbury=457,
width=15.0cm}
\caption{\label{fig:pla_eff} Absolute efficiency of the 30 \% HPGe 
detector: The data points are from the measurements with the
calibration sources, the drawn line is an interpolation, which has
been fitted to the data points with a least squares fit.  Note the
doubly logarithmic scale.  }
\end{figure}

Because the activity of the platinum samples was relatively low, they
have been mounted directly in front of the 30 \% HPGe detector used
for the activation measurement.  Since the activity of the available
calibration sources was much higher, only a small number of lines
could be used for the calibration measurements of the HPGe
detector. However, some of the calculated data points for the
efficiency still have relatively big errors, as is shown in
Fig.~\ref{fig:pla_eff}. These errors are mainly due to summing
effects resulting from the high count rate of the detector.  For the
calculations we used a doubly logarithmic interpolation of the data
points in the relevant energy interval.

It had to be considered that due to the thickness of the platinum
disks of 0.125 mm a portion of the emitted $\gamma$ rays was
absorbed within the disks. This portion has been
estimated using the computer code GEANT. The necessary
corrections were of the order of a few percent, except for the 77.4
keV line of the $^{197}$Pt, where approximately two thirds of the
$\gamma$ rays were absorbed within the platinum sample.

\section{Experimental results}
\label{sec:res}

\subsection{Results for the cross section $\sigma_{(\gamma,n)}$}
\label{subsec:sigma0}
The cross section parameters $\sigma_0$ from Eq.~(\ref{eq:sqrt}) have
been calculated twice for every isotope,
using both the spline interpolated and the by the factor $F$ 
corrected photon spectra. 
The two results are given in Tab.~\ref{tab:summ}, averaged over all
measurements with different endpoint energies.

For $^{190}$Pt an additional measurement has been performed.
This was necessary because the evaluation of the $^{190}$Pt lines
was close to the statistical limit. 
For this measurement the platinum target was placed
directly  behind the radiator target because of the much higher photon
flux. 
In this target position the ($\gamma$,$\gamma'$) lines of the
$^{11}$B samples could not be measured. Therefore, the shape of the
photon spectrum could not be determined. 
However, since the neutron separation energies of $^{190}$Pt and of
$^{192}$Pt are close to each other (Tab.~\ref{tab:summ}), we were able
to calculate the ratio of the cross sections of these isotopes by
determining the ratio of the number of counts in their respective
lines. Then we calculated the cross section of $^{190}$Pt by
multiplying this ratio with the known value for the cross section of
$^{192}$Pt from the previous measurement. This value is considered in
the average in Tab.~\ref{tab:summ}.

For the calculation of errors, the uncertainties in the detector
efficiencies, the statistical errors from the number of counts in the
respective peaks and the errors in the relative and absolute
intensities of the decay
lines have been considered. In order to estimate the error 
resulting from the uncertainty in the shape of the photon spectra, we
calculated the mean difference of the interpolated spectra and the
spectra which were adjusted by the correction factor $F$. 
For $^{190}$Pt an additional error of 7\% has been considered,
resulting  from an uncertainty in the natural abundance of this isotope.

\subsection{Results for the $(\gamma,n)$ reaction rates}
\label{sec:gnrates}

The ($\gamma$,n) reaction rates $\lambda$ have been determined from
$\sigma_0$ and with the superposition method 
(see Sect.~\ref{sec:analysis}).
With both methods, the reaction rates in the 
complete relevant temperature region ($T_9=2-3$) have been calculated. 
The results are given in Tables \ref{tab:lambda192}, \ref{tab:lambda198},
and \ref{tab:lambda190}.  These results are averaged over all
evaluated lines of the respective isotopes.  For the superposition of
the bremsstrahlung spectra, the bremsstrahlung spectra adjusted by the
correction factor $F$ of Eq.~(\ref{eq:faktor}) have been used.  For the
conventional method, only the values of $\sigma_0$ calculated with the
corrected photon spectra have been used in order to be able to compare
the results of both methods.  For the isotope $^{190}$Pt the
superposition method could not be used because the decay lines of 
$^{189}$Pt could not be evaluated in all measurements. For this isotope
the average value of $\sigma_0$ which is given in Tab.~\ref{tab:summ}
has been used for the calculation in the conventional method.  The
relative errors for the reaction rates calculated by the conventional
method are the same as the error for the parameter $\sigma_0$.

\begin{figure}
\epsfig{figure=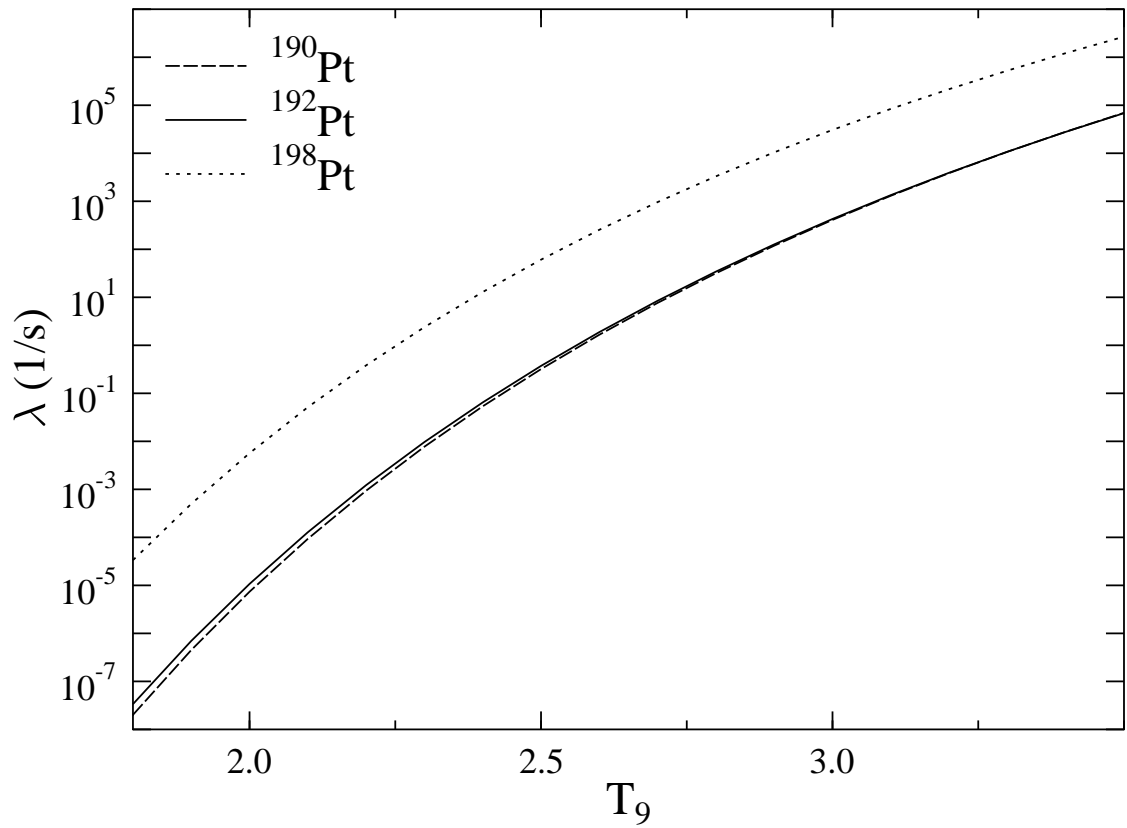,bbllx=5,bblly=268,bburx=502,bbury=637,
width=15.0cm}
\caption{ \label{fig:tempverlauf} Temperature dependence of the
($\gamma$,n) reaction rates $\lambda$ in the complete relevant
temperature region from T$_9 = 2-3$. The values have been calculated
by the superposition of the different bremsstrahlung spectra.  T$_9$
is the temperature in 10$^9$ K.  
Note the much larger reaction rate of $^{198}$Pt because of the
significantly smaller neutron separation energy.
}
\end{figure}

The temperature dependence of the reaction rates is shown in
Fig.~\ref{fig:tempverlauf} for the isotopes $^{190}$Pt, $^{192}$Pt, and
$^{198}$Pt. The given values are calculated using the superposition
method.  The comparison of the results of both methods shows good
overall agreement. The error of the conventional method seems to be
smaller than the error of the superposition method because of the
additional error resulting from the difference between thermal and
superposed photon spectra. It has to be pointed out that the most
important systematic uncertainty in the conventional method cannot 
be estimated,
i.e.\ the uncertainty how good Eq.~(\ref{eq:sqrt}) approximates the
shape of the cross section energy dependence. Therefore, the
superposition method should prove much more reliable.

\subsection{Comparison with other results}
\label{subsec:other}
\begin{figure}
\epsfig{figure=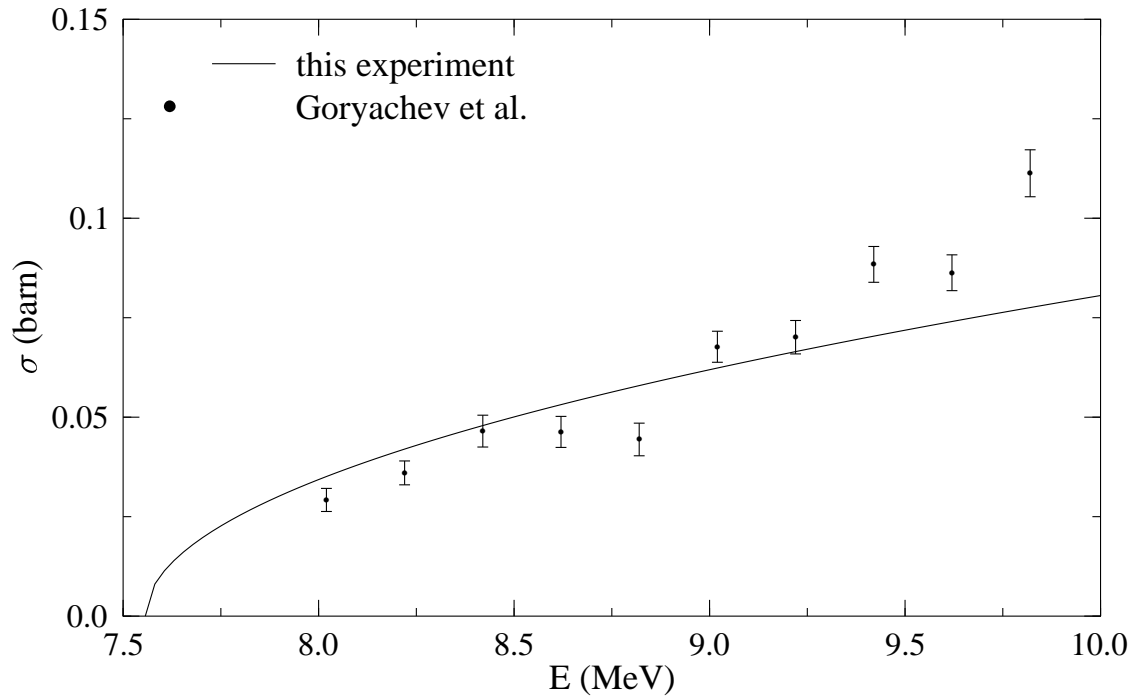,bbllx=16,bblly=254,bburx=605,bbury=616,
width=15.0cm}
\caption{ \label{fig:goryachev}Comparison of our results for the
($\gamma$,n) cross section of $^{198}$Pt to a previous experiment by
Goryachev {\it et al.}~\protect\cite{Gor78}.}
\end{figure}

In Fig.~\ref{fig:goryachev} our result for the ($\gamma$,n) cross
section of the isotope $^{198}$Pt is compared with a previous
experiment by Goryachev {\it et al.} \cite{Gor78}. They measured the
($\gamma$,n) cross section at various energies in and below the GDR
region whereas we determined the cross section 
parameter $\sigma_0$. Therefore
a comparison of both results is only possible if one assumes the
behavior of the cross section to follow Eq.~(\ref{eq:sqrt}).  Since
this parametrization only holds in the vicinity of the threshold
energy, the deviations at higher energies are not surprising.

Previous direct experiments had to be performed with huge amounts of
highly enriched target material of the order of several grams. Such an
amount of highly enriched material is not available for the low
abundant platinum isotopes $^{190}$Pt and $^{192}$Pt, and hence
no data exist in literature for these isotopes and for the p isotopes
of other elements \cite{Die88,CDFE}. To our knowledge our data are the
first experimental ($\gamma$,n) cross sections and 
astrophysical reaction rates measured for the p-nuclei.

\section{Theoretical consideratons}
\label{sec:theo}

\subsection{Calculation of laboratory and stellar rates}
\label{subsec:calc}
Experimental data can only provide photodisintegration rates of targets in
their ground state. In a stellar plasma of the required temperature and
density, nuclei are thermally equilibrated with their environment and
therefore also excited states will be populated. The laboratory cross section
$\sigma_{(\gamma,{\rm n})}^{\rm lab}=\sum _{\nu}\sigma^{0\nu }$ has to be
replaced by the stellar cross section
\begin{equation}
\label{csstar}
\sigma ^{*}(E,T)={\sum _{\mu }(2J^{\mu }+1)\exp (-E^{\mu }/kT)
\sum _{\nu }\sigma ^{\mu \nu }(E)\over 
\sum _{\mu }(2J^{\mu }+1)\exp (-E^{\mu }/kT)}\quad ,
\end{equation}
where $\mu$ and $\nu$ denote the target states and the states in the final
nucleus, respectively. Depending on plasma temperature $T$, spins $J^{\mu}$ and
location $E^{\mu}$ of the target states, the stellar cross section can become
considerably different from the one measured in the laboratory.

Up to now, astrophysical photodisintegration rates were calculated in a purely
theoretical way by deriving them from capture rates via detailed balance
(e.g.\ \cite{Fow67,hwf76,rath}). It should be noted that only {\it stellar}
capture rates can be used to correctly apply detailed balance and, vice versa,
only stellar photodisintegration rates can be used for the derivation of
the respective stellar capture rates. Those stellar capture rates are usually
also calculated theoretically. Detailed balance is valid for exoergic
reactions ($Q>0$) and it has recently been shown that it is still
quite accurate for charged particle capture with $Q<0$ \cite{RTO95}.

Nevertheless, laboratory measurements are an important way to check the
validity of the involved assumptions and the nuclear properties needed
for the prediction of cross sections and rates. A direct measurement of the
($\gamma$,n) cross section can not only test detailed balance but
also the description used for the low-energy tail of the GDR and the
neutron optical potential for nuclei in excited states. Measurements using
neutron capture are more limited in this respect.

Here, we compare the experimental results to calculations with the
statistical model code NON-SMOKER \cite{rath}.
For the calculations presented here, 
the code NON-SMOKER was modified in such a way to be able to calculate 
photodisintegration reactions directly instead of deriving them via
detailed balance (like e.g. in Ref.\ \cite{rath}). This modification also
allows to compute rates for targets in the ground state which can easily be
compared to our experimental results. The theoretical results are collected
in Tab.\ \ref{tab:theory} and a comparison to the data is shown in Tab.\ 
\ref{tab:theoryvgl}.

\subsection{Discussion}
\label{subsec:disc}
In general, the theoretical rates are in good agreement with the present
data, as can be seen from Tab.\ \ref{tab:theoryvgl}. For the global
statistical model calculations compared here, a typical average
deviation of the order of 30 \% should be expected but locally larger
deviations up to a factor of two are possible. In this sense, the agreement
is excellent for $^{198}$Pt and acceptable for the other isotopes, depending
on which error is used. A visible trend of the accuracy with mass cannot be
established.

The temperature dependence of the ratios from Tab.\ \ref{tab:theoryvgl} can,
in principle, be used to study how well the energy dependence of the cross
section is reproduced. The latter, in turn, is mainly determined by the
low-energy tail of the GDR. In addition, there is a weaker dependence on the
neutron potential in the exit channel. The effect of possible inaccuracies 
in the theoretical description of those properties are weakened, however,
by the smoothing due to the integration over the effective Gamow-like
energy window in the derivation of the rate. A direct comparison of
cross sections may be more sensitive but the relevant quantity for
astrophysics is the reaction rate.

As discussed before (Sec.\ \ref{sec:gnrates}), the values derived by
the use of the conventional method agree well with those of the
superposition method, indicating that the threshold behavior of the
($\gamma$,n) cross section is roughly reproduced by Eq.\
(\ref{eq:sqrt}). This is supported by the comparison to the
($\gamma$,n) cross sections of Ref.\ \cite{Gor78}.  However, it cannot
be expected in general that the cross section exhibits a structureless
behavior proportional to $\sqrt{E-E_{\rm{thr}}}$.  The astrophysically
relevant cross section depends on the dipole strength distribution in
a narrow region above the threshold.  Therefore, it is of importance
to study this strength distribution experimentally. It is well known
that the E1 strength exhibits significant fine structure like the
so-called `pygmy resonance' close to the particle threshold
\cite{Hart00,Kop90,Iga86,End00}.  The presence or absence of such a strength
accumulation can change the astrophysical reaction rates
significantly, while it will affect the position and width of the
Gamow-like window on a small scale only.

There seem to be minor deviations between the two methods at $T_9=2.0$
for $^{198}$Pt and at $T_9=3.0$ for $^{192}$Pt. These minor
discrepancies can be explained by the fact that the assumed threshold
behavior of the cross section underestimates the cross section at
higher energies (see also Fig.~\ref{fig:goryachev}) which leads to a
somewhat smaller slope of the reaction rate in the conventional analysis.

Within the errors the
temperature dependence of the theoretical rates agrees with the data.
Oddly enough, there seems to be a slight difference for $^{198}$Pt which is
otherwise reproduced best. However, further conclusions can only be drawn
pending a reduction of the experimental error bars.

Since the theoretical ($\gamma$,n) values have been confirmed for
laboratory rates, detailed
balance can be tested by comparing our stellar rates to the stellar 
photodisintegration rates derived from stellar neutron capture rates,
as provided in Refs.\ \cite{rath,bao}. We find perfect agreement in all
cases, confirming the validity of detailed balance.

As can be seen from Tab.\ \ref{tab:theory}, the stellar rates are larger
by several orders of magnitude than the laboratory rates due to the
facilitation of photodisintegration for thermally excited targets. This effect
is sensitive to the used level density and structure but cannot be directly tested in the
laboratory.

\section{Summary and conclusions}
\label{sec:summ}
It has been shown that a thermal distribution of photons at
astrophysically relevant temperatures can be simulated by the
appropriate superposition of several bremsstrahlung spectra with
different endpoint energies. Several platinum samples were irradiated
with this quasi-thermal photon spectrum, and the ($\gamma$,n) reaction
of $^{190}$Pt, $^{192}$Pt, and $^{198}$Pt was analyzed using the
photoactivation technique. The high sensitivity of this method allows
the measurement of the ($\gamma$,n) reaction even for isotopes with
very low natural abundances.

The measured reaction rates in the laboratory have been compared to a
statistical model calculation, and good agreement was found for all
analyzed isotopes. Furthermore, stellar reaction rates have been
calculated which are enhanced by the thermal population of excited
states in the target nucleus.

These photon induced reactions are important for the nucleosynthesis
of the neutron-deficient p-nuclei which are synthesized in the
astrophysical $\gamma$ process in supernova explosions at temperatures
of $T_9 = 2 - 3$. Almost no experimental data exist for
the reactions relevant for the $\gamma$ process. For a better
understanding of the $\gamma$ process more experimental data for
($\gamma$,n) and ($\gamma$,$\alpha$) reactions in the astrophysically
relevant energy region are required. A systematic study is necessary
to verify the predictions of statistical model calculations.

\acknowledgements We thank the S--DALINAC group around H.-D.~Gr\"af
for the reliable beam during the photoactivation and
U.~Kneissl  and A.~Richter for valuable discussions.  
This work was supported by the
Deutsche Forschungsgemeinschaft (contracts Zi\,510/2-1 and
FOR\,272/2-1). T.\ R.\ is supported by a PROFIL professorship from the
Swiss National Science Foundation (grants 2124-055832.98,
2000-061822.00) and by the NSF (grant NSF-AST-97-31569).

\end{multicols}

%
%

\begin{table}
\caption{\label{tab:platin} Properties of the examined platinum
isotopes and the decay $\gamma$ rays following the electron capture or
$\beta^-$ decays. The absolute intensities have been taken from
\protect\cite{NDS}. Note the big errors in the absolute intensities,
resulting from a large uncertainty in the conversion factor from
relative to absolute intensities in\protect\cite{NDS}.}
\begin{center}
\begin{tabular}{cccccc} 
reaction & $E_{\rm{thr}}$ & daughter  & decay  &
energy & intensity   \\
 & (keV) & & & (keV) & per decay \\ \hline
$^{190}$Pt($\gamma$,n)$^{189}$Pt & 8911 & $^{189}$Ir   
	& $\epsilon$ & 568.9         & 0.071 $\pm$ 0.006 \\  
        &  &              &  & 608.0         & 0.081  $\pm$ 0.021  \\ \hline
$^{192}$Pt($\gamma$,n)$^{191}$Pt & 8676 & $^{191}$Ir  
	& $\epsilon$ & 360.0         & 0.06  $\pm$ 0.015 \\ 
        &  &              &  & 409.4         & 0.08  $\pm$ 0.021 \\ 
        &  &              &  & 538.9         & 0.137  $\pm$ 0.035 \\ \hline
$^{198}$Pt($\gamma$,n)$^{197}$Pt & 7557 & $^{197}$Au  
	& $\beta^-$  & 77.4          & 0.172   $\pm$ 0.025 \\ 
        &  &              &  & 191.4        & 0.037  $\pm$ 0.004 \\ 
        &  &              &  & 268.8        & 0.0023  $\pm$ 0.0003

\end{tabular}
\end{center}
\end{table}

\begin{table}
\caption{\label{tab:summ}Results for the cross section
parameter $\sigma_0$. The values are weighted averages over all
measured lines of the respective isotope. The values in the last
column are weighted averages of the first two, except for
$^{190}$Pt, where the result of an additional measurement has been
included (details see Sec.\ \ref{subsec:sigma0}).}
\begin{center}
\begin{tabular}{cccc}
 isotope &  $\sigma_0$ (mb) & $\sigma_0$ (mb) & $\sigma_0$ (mb) \\
 & interpolation & correction factor & average \\ \hline
$^{190}$Pt  & 488 $\pm$ 146 & 518 $\pm$ 153 & 300 $\pm$ 100\tablenotemark[1]\\ 
$^{192}$Pt  & 118 $\pm$ 34  & 103 $\pm$ 30 & 111 $\pm$ 32 \\  
$^{198}$Pt  & 161 $\pm$ 27  & 142 $\pm$ 24 & 152 $\pm$ 25     
\end{tabular}
{\tablenotetext[1]{including additional measurement (see Sec.\ \ref{subsec:sigma0}).}}
\end{center}
\end{table}

\begin{table}
\caption{\label{tab:bordaten}Properties of the $^{11}$B levels which
were used for the determination of the shape of the bremsstrahlung
spectra by the $^{11}$B($\gamma$,$\gamma'$) measurement. The decay
widths $\Gamma$ were taken from the compilation \protect\cite{aizen}. $I$ is
the energy-integrated cross section.}
\begin{center}
\begin{tabular}{cccc} 
$E_x$    & $\Gamma_0 / \Gamma$ 
&  $\Gamma$  & $I$  \\
 (keV) & & (eV) & (10$^3$ eVfm$^2$)  \\ \hline
2124.7                   & 1                         & 0.12   & 5.1
$\pm$ 0.4 \\  
4443.9                   & 1                         & 0.56   & 16.3
$\pm$ 0.5\\  
5019.1                   & 0.856                     & 1.963  & 21.9
$\pm$ 0.8\\  
7282.9                   & 0.87                      & 1.149  & 9.5
$\pm$ 0.7\\  
8916.3                   & 0.95                      & 4.368  & 28.6
$\pm$ 1.4  
\end{tabular}
\end{center}
\end{table}

\begin{table}
\caption{\label{tab:lambda192} Average weighted values of the
($\gamma$,n) reaction rate $\lambda$ for the nucleus $^{192}$Pt in the
complete temperature range relevant for the astrophysical $\gamma$ 
process.  }
\begin{center}
\begin{tabular}{ccc} 
 temperature & $\lambda$ conventional\tablenotemark[1]\tablenotemark[2] &
 $\lambda$ superposition\tablenotemark[1] \\
 (10$^9$ K) &  (s$^{-1}$) &  (s$^{-1}$) \\ \hline

  2.0 & (9.6 $\pm$ 0.7) $\times$ 10$^{-6}$ & (10.7 $\pm$ 2.7) $\times$
  10$^{-6}$ \\ 
  2.1 & (1.14 $\pm$ 0.09) $\times$ 10$^{-4}$ & (1.28 $\pm$ 0.29)
  $\times$ 10$^{-4}$ \\ 
  2.2 & (1.09 $\pm$ 0.08) $\times$ 10$^{-3}$ & (1.22 $\pm$ 0.25) $\times$
  2.10$^{-3}$ \\
  2.3 & (8.50 $\pm$ 0.63) $\times$ 10$^{-3}$ & (9.67 $\pm$ 1.83)
  $\times$ 10$^{-3}$ \\
  2.4 & (5.62 $\pm$ 0.42) $\times$ 10$^{-2}$ & (6.46 $\pm$ 1.13)
  $\times$ 10$^{-2}$ \\ 
  2.5 & 0.320 $\pm$ 0.024 & 0.372 $\pm$ 0.060 \\
  2.6 & 1.60 $\pm$ 0.12 & 1.88 $\pm$ 0.29 \\
  2.7 & 7.11 $\pm$ 0.53 & 8.47 $\pm$ 1.22 \\
  2.8 & 28.4 $\pm$ 2.1 & 34.3 $\pm$ 4.7 \\
  2.9 & 104 $\pm$ 8 & 126 $\pm$ 17 \\
  3.0 & 346 $\pm$ 26 & 428 $\pm$ 55
\end{tabular}
{\tablenotetext[1]{ the calculated values of $\lambda$ are
subject to an additional error of 27.8\% resulting from the
uncertainty in the absolute intensity of the respective $^{191}$Pt lines
and the uncertainty in the efficiency of the detectors used for the
($\gamma,\gamma'$) measurement.}}
{\tablenotetext[2]{an additional error resulting from the
approximation of the shape of the cross section taken from
Eq.~(\ref{eq:sqrt}) cannot be estimated.}}
\end{center}
\end{table}

\begin{table}
\caption{\label{tab:lambda198} Average weighted values of the
($\gamma$,n) reaction rate $\lambda$ for the nucleus $^{198}$Pt in the
complete temperature range relevant for the astrophysical $\gamma$ 
process.  }
\begin{center}
\begin{tabular}{ccc} 
 temperature  &  $\lambda$ conventional\tablenotemark[1]\tablenotemark[2] 
 & $\lambda$ superposition\tablenotemark[1] \\
 (10$^9$ K) & (s$^{-1}$) & (s$^{-1}$)  \\   \hline 
 
  2.0 & (7.19 $\pm$ 0.25) $\times$ 10$^{-3}$ 
  & (5.68 $\pm$ 1.00) $\times$ 10$^{-3}$ \\ 
  2.1 & (6.26 $\pm$ 0.22) $\times$ 10$^{-2}$
  & (5.14 $\pm$ 0.84) $\times$ 10$^{-2}$ \\ 
  2.2 & 0.449 $\pm$ 0.016 & 0.383 $\pm$ 0.057 \\ 
  2.3 & 2.73 $\pm$ 0.10 & 2.39 $\pm$ 0.33 \\ 
  2.4 & 14.3 $\pm$ 0.5 & 12.9 $\pm$ 1.7 \\ 
  2.5 & 65.6 $\pm$ 2.3 & 60.8 $\pm$ 7.3 \\ 
  2.6 & 269 $\pm$ 10 & 255 $\pm$ 29 \\
  2.7 & 996 $\pm$ 35 & 961 $\pm$ 102 \\ 
  2.8 & (3.37 $\pm$ 0.12) $\times$ 10$^{3}$ 
  & (3.31 $\pm$ 0.33) $\times$10$^{3}$ \\
  2.9 & (1.05 $\pm$ 0.04) $\times$ 10$^{4}$ 
  & (1.05 $\pm$ 0.1) $\times$ 10$^{4}$ \\ 
  3.0 & (3.03 $\pm$ 0.11) $\times$10$^{4}$ 
  & (3.07 $\pm$ 0.27) $\times$10$^{4}$
\end{tabular}
{\tablenotetext[1]{the calculated values of $\lambda$ are
subject to an additional error of 16.3\%  resulting from the
uncertainty in the absolute intensity of the respective $^{197}$Pt lines
and the uncertainty in the efficiency of the detectors used for the
($\gamma,\gamma'$) measurement.}}
{\tablenotetext[2]{an additional error resulting from the
approximation of the shape of the cross section taken from
Eq.~(\ref{eq:sqrt}) cannot be estimated.}}
\end{center}
\end{table}

\begin{table}
\caption{\label{tab:lambda190} Average weighted values of the
($\gamma$,n) reaction rate $\lambda$ for the nucleus $^{190}$Pt in the
complete temperature range relevant for the astrophysical $\gamma$ 
process. Calculated using the conventional method.}
\begin{center}
\begin{tabular}{cc} 
 temperature  &  $\lambda$\tablenotemark[1]\tablenotemark[2] \\
 (10$^9$ K) &  (s$^{-1}$) \\ \hline 
  2.0  & 7.38 $\times$ 10$^{-6}$  \\ 
  2.1  & 9.32 $\times$ 10$^{-5}$  \\ 
  2.2  & 9.37 $\times$ 10$^{-4}$  \\  
  2.3  & 7.73 $\times$ 10$^{-3}$  \\ 
  2.4  & 5.36 $\times$ 10$^{-2}$  \\  
  2.5  & 0.319 \\  
  2.6  & 1.66  \\  
  2.7  & 7.65  \\ 
  2.8  & 31.7 \\ 
  2.9  & 119  \\  
  3.0  & 409   
\end{tabular}
{\tablenotetext[1]{the calculated values of $\lambda$ are
subject to an error of 33\%.}}
{\tablenotetext[2]{an additional error resulting from the
approximation of the shape of the cross section taken from
Eq.~(\ref{eq:sqrt}) cannot be estimated.}}
\end{center}
\end{table}

\begin{table}
\caption{\label{tab:theory} Theoretical ($\gamma$,n) reaction rates 
in 1/s for the target in the ground state $\lambda^{\rm gs}$ and a
thermally excited target $\lambda^*$, calculated with the NON-SMOKER code.}
\begin{center}
\begin{tabular}{cllllll}
\multicolumn{1}{c}{Temperature}&\multicolumn{2}{c}{$^{190}$Pt}&
\multicolumn{2}{c}{$^{192}$Pt}&\multicolumn{2}{c}{$^{198}$Pt}\\
\multicolumn{1}{c}{10$^9$ K}&\multicolumn{1}{c}{$\lambda^{\rm gs}$}&
\multicolumn{1}{c}{$\lambda^*$}&\multicolumn{1}{c}{$\lambda^{\rm gs}$}&
\multicolumn{1}{c}{$\lambda^*$}&\multicolumn{1}{c}{$\lambda^{\rm gs}$}&
\multicolumn{1}{c}{$\lambda^*$}\\ \hline
2.0&4.11 $\times$ 10$^{-6}$&1.70 $\times$ 10$^{-2}$&1.75 $\times$ 10$^{-5}$&4.58 $\times$ 10$^{-2}$&6.85 $\times$ 10$^{-3}$&1.66 $\times$ 10$^{0}$\\
2.5&1.80 $\times$ 10$^{-1}$&7.47 $\times$ 10$^2$&5.71 $\times$ 10$^{-1}$&1.41 $\times$ 10$^3$&5.79 $\times$ 10$^{1}$&1.49 $\times$ 10$^{4}$\\
3.0&2.37 $\times$ 10$^2$&9.23 $\times$ 10$^5$&6.16 $\times$ 10$^2$&1.35 $\times$ 10$^6$&2.52 $\times$ 10$^{4}$&6.32 $\times$ 10$^{6}$
\end{tabular}
\end{center}
\end{table}

\begin{table}
\caption{\label{tab:theoryvgl} Ranges of the ratio $\lambda^{\rm gs}/\lambda^{\rm exp}$ determined by the quoted errors on the experimental rates. Except for
$^{190}$Pt, the values extracted by the superposition method are used.}
\begin{center}
\begin{tabular}{clll}
\multicolumn{1}{c}{Temperature}&\multicolumn{1}{c}{$^{190}$Pt}&
\multicolumn{1}{c}{$^{192}$Pt}&\multicolumn{1}{c}{$^{198}$Pt}\\
\multicolumn{1}{c}{10$^9$ K}&\multicolumn{1}{c}{$\lambda^{\rm gs}/\lambda^{\rm exp}$}&
\multicolumn{1}{c}{$\lambda^{\rm gs}/\lambda^{\rm exp}$}&
\multicolumn{1}{c}{$\lambda^{\rm gs}/\lambda^{\rm exp}$}\\ \hline
2.0&$0.42-0.83$&$1.31-2.19$&$1.03-1.46$\\
2.5&$0.42-0.84$&$1.32-1.83$&$0.85-1.08$\\
3.0&$0.44-0.87$&$1.28-1.65$&$0.75-0.90$\\
\end{tabular}
\end{center}
\end{table}

\end{document}